\documentclass{article}
\usepackage{spconf,amsmath,graphicx,url}

\title{STAN: A Stuttering Therapy Analysis Helper} 

\name{Sebastian P. Bayerl$^1$, Marc Wenninger$^2$,  J. Schmidt$^2$, Alexander Wolff von Gudenberg$^3$, Korbinian Riedhammer$^1$}
\address{
  $^1$Technische Hochschule Nürnberg Georg Simon Ohm, 
  $^2$Technische Hochschule  Rosenheim \\ 
  $^3$Institut der Kasseler Stottertherapie}

\begin{document}

\maketitle

\begin{abstract}
\vspace{-1mm}
Stuttering is a complex speech disorder identified by repetitions, prolongations of sounds, syllables or words and blocks while speaking.
Specific stuttering behaviour differs strongly, thus needing personalized therapy. 
Therapy sessions require a high level of concentration by the therapist.
We introduce STAN, a system to aid speech therapists in stuttering therapy sessions.
Such an automated feedback system can lower the cognitive load on the therapist and thereby enable a more consistent therapy as well as allowing analysis of stuttering over the span of multiple therapy sessions.
\footnote{This work is supported by the Bayerisches Staatsministerium für Bildung und Kultus, Wissenschaft und Kunst and BayWISS.}

\end{abstract}
\vspace{-1mm}
\section{Introduction}
\vspace{-1mm}
Stuttering is a complex speech disorder which affects about 1\,\% of the population \cite{Carlson2012}.
The disorder is caused by problems in nerve coordination between areas in the brain.
It can be identified by repetitions, prolongations of sounds, syllables or words, and blocks while speaking.
The condition is generally treatable but not curable.
Only little research exists on the topic of automated assessment of stuttering and stuttering therapy. 
As the symptoms of this condition are different for each person a generalized solution is difficult to achieve.
The individual symptoms of each person who stutters (PWS) depend on factors such as the communication situation, the linguistic complexity of an utterance and the typical phased progress of the speech disorder \cite{handbook_stuttering_2008,packman2004theoretical}.

A standard speech recognition system was used to evaluate vowel and fricative duration on a standard reading task to differentiate a PWS and normal speakers in \cite{noth_stuttering_2000}.
\cite{mfcc_stuttering_chee_2009} is focused on the classification of prolongations and repetitions by extracting Mel Frequency Cepstral Coefficients (MFCC) and then train k-NN and LDA classifiers on a small sample taken from the University College London Archive of Stuttered Speech\footnote{Available at \urlstyle{rm}\url{https://www.uclass.psychol.ucl.ac.uk/uclassfsf.htm}}.
Swietlicka et al.\ use neural networks to discriminate between syllable repetitions, blocks before words that start with a plosive, and phone prolongations \cite{ann_stuttering_swietlicka_2013}.
A Text-based approach focused on the detection of repetitions was proposed in \cite{lightly_supervised_stuttering_alharbi_2018}.
Evaluations on a stuttering therapy based dataset have shown that evaluating phone durations can help identify sections of stuttered speech \cite{bayerl_2020}.
The approach in \cite{ochi_soft_articulartory_2018} is of special interest because of its focus on the automatic evaluation of soft articulatory contact, a technique commonly taught in stuttering therapy.
Detecting these is important to evaluate PWS that already underwent speech therapy.

To assess the severity of stuttering and eventual therapy success it is important to measure stuttering reliably. 
A typical stuttering diagnosis consists of the objective and subjective evaluation of the stuttering symptoms as well as the evaluation of the impact on everyday life.
Such an evaluation is usually performed by a speech therapist.
The cognitive load of such a task is high as it involves leading a conversation with the PWS while keeping track of disfluencies and counting syllables. 
Evaluating recordings afterwards is often too time consuming, but a therapy sessions could greatly benefit from automated analysis.
Introducing automated annotations to highlight possible stuttering events increases the time that can be spent discussing problematic sections with the client.

To support this tasks we introduce \textbf{STAN} (Stuttering Therapy Analysis Helper).
STAN can help identify hotspots and assist therapists in their assessment by offering additional information \cite{bayerl_2020, noth_stuttering_2000, jiao_phono_2017}.

\vspace{-1mm}
\vspace{-1mm}
\section{System and Demo description}
\vspace{-1mm}
\paragraph*{Analysis}
\begin{figure*}[!ht]
    \centering
    \includegraphics[width=
    \textwidth]{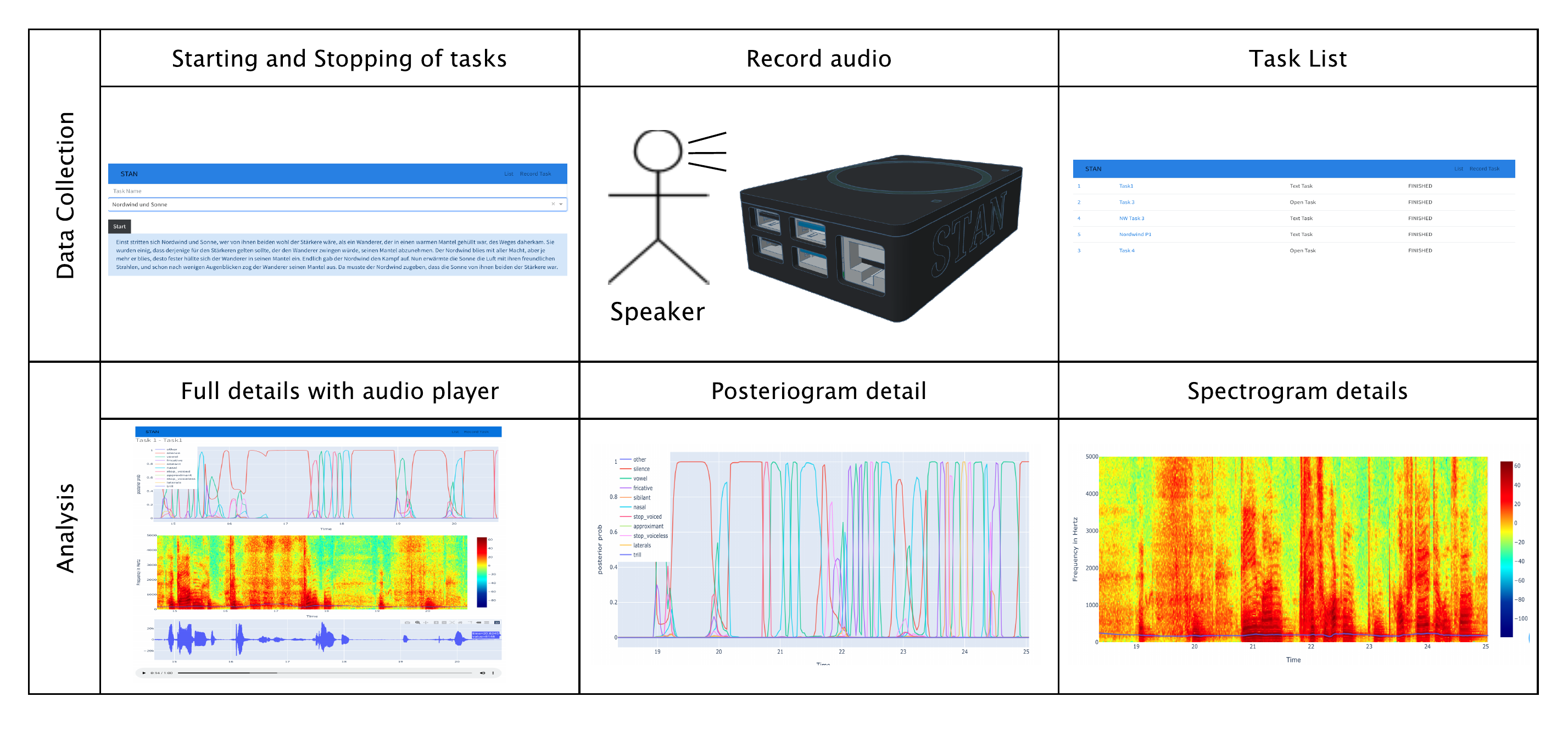}
    \vspace{-3mm}
    \caption{Depiction of STANs data collection and analysis interface}
    \label{fig:stan_flowchart}
\end{figure*}
\vspace{-3mm}

STAN is based on a headless Raspberry Pi 4 running Raspbian 10 and a ReSpeaker 4-Mic Array for recording. 
The recordings and analysis results will only be stored on device to preserve privacy. 
The system can be operated from a second device using a Web-UI.

\vspace{-1mm}
\paragraph*{Data Collection}

The system provides support for two types of task: 
a standardized reading task and
open client therapist conversation.
The open client therapist conversation just listens in, while the standardized reading task displays text upon starting the task and records audio along with it.
After finishing the tasks the recording is stored on device and the feature extraction and analysis pipeline l runs.

A background task extracts MFCC features in both cases and performs energy-based voice activity detection and segmentation.
Furthermore, phonological posteriors are extracted by making a forward pass through the acoustic model of a kaldi-based ASR system.
A phone-based speech recognition is run to extract phonetic text.
For the voiced segments the pitch is also being extracted.
The results are stored in a database on device.

The analysis dialogue consists of interactive plots that allow zooming and panning.
One plot contains phonological posteriors as well as a pane with a spectrogram and the plotted pitch information. 
The events marked as potential stutters are added to the plotted phonological posteriors.
All plots are interlinked to provide seamless navigation.
The spectrogram is only displayed for selected timespans below ten seconds.

\vspace{-1mm}
\paragraph*{Demo} The demonstration will contain a short dialogue between a therapist and a PWS. 
STAN will be used to record the session and after processing the recorded audio give feedback in a web interface.
The web interface gives an interactive overview of the recording, containing an audio player, a spectrogram view, as well as phonological posteriors to provide the therapist with explainable feedback to help better understand the classification results. 
Phonological posteriors are acquired by performing a forward pass of the acoustic model of an ASR system and the phones are then mapped to the phonological class they belong to.
Those are comprehensible for the speech therapists \cite{jiao_phono_2017}.
To only analyze the relevant speech segments (PWS), the therapist's speech segments are removed using an X-Vector and SVM-based speaker id system.







The system introduced here is a mobile, low-cost solution that is based entirely on open-source components and cost-effective hardware. 
We provide an effective solution that will be evaluated in the field in cooperation with a large German stuttering therapy provider. 


\bibliographystyle{IEEEbib}

\footnotesize{\bibliography{mybib}}

\end{document}